\documentclass[aps,pre,reprint,onecolumn,superscriptaddress,11pt]{revtex4-2}

\usepackage{amsmath,amssymb,xcolor,upgreek,layouts, orcidlink,siunitx,mathtools} 
\usepackage{graphicx} 
\usepackage{booktabs} 
\usepackage[a4paper,margin=1in]{geometry}
\usepackage{orcidlink}

\newcommand{\sgn}{\mathrm{sgn}}
\newcommand{\blue}[1]{{#1}} 

\begin{document}
\title{Swimming-limited aggregation of bacteria in liquid crystals}

\author{Guillaume Sintès} 
\affiliation{Laboratoire de Physique et Mécanique des Milieux Hétérogènes (PMMH), ESPCI Paris, PSL University, CNRS, Sorbonne University, and Paris Cité University, 75005 Paris, France}
\affiliation{Laboratoire Gulliver, ESPCI Paris, PSL University, CNRS, 75005 Paris, France}

\author{Martyna Goral\,\orcidlink{0000-0001-6194-015X}}
\affiliation{Laboratoire de Physique et Mécanique des Milieux Hétérogènes (PMMH), ESPCI Paris, PSL University, CNRS, Sorbonne University, and Paris Cité University, 75005 Paris, France}
\affiliation{Laboratoire Gulliver, ESPCI Paris, PSL University, CNRS, 75005 Paris, France}

\author{Teresa L\'{o}pez-Le\'{o}n\,\orcidlink{0000-0002-3272-6389}}
\affiliation{Laboratoire Gulliver, ESPCI Paris, PSL University, CNRS, 75005 Paris, France}

\author{Anke Lindner\,\orcidlink{0000-0002-5007-9568}}
\email[Corresponding author: ]{anke.lindner@espci.fr}
\affiliation{Laboratoire de Physique et Mécanique des Milieux Hétérogènes (PMMH), ESPCI Paris, PSL University, CNRS, Sorbonne University, and Paris Cité University, 75005 Paris, France}
\affiliation{Institut Universitaire de France (IUF), Paris, France}

\author{Maria Tătulea-Codrean\,\orcidlink{0000-0001-5518-7958}}
\email[Corresponding author: ]{m.tatulea-codrean@uva.nl}
\affiliation{Institute of Physics, University of Amsterdam, 1098 XH Amsterdam, The Netherlands} 

\date{\today}

\begin{abstract}

Aggregation and fragmentation processes are widespread in engineering and the natural world. Here, we investigate a distinct colloidal aggregation mechanism in an active system of motile bacteria in highly anisotropic environments. Specifically, we examine \textit{Escherichia coli} bacteria swimming in one-dimensional confinement within nematic liquid crystals and observe long-lived chains of bacteria swimming along the nematic director. Crucially, we find that longer chains swim faster, in apparent contradiction to fundamental force-balance models that predict the swimming speed to be independent of chain length, as chains should swim at the average speed of their individual components. The seeming discrepancy is resolved by recognizing that chains do not form randomly but self-organize due to the relative velocities between bacteria.
To elucidate the physical mechanism behind this active aggregation process, we combine our experimental findings with a minimal model of nearest-neighbour aggregation and agent-based simulations of active particles aggregating in one dimension. Consistent with experimental observations, our agent-based simulations reveal a positive correlation between the length and speed of dynamically self-assembled chains of active particles,  with the correlation depending on the variance of the individual speed distribution and diminishing over time. 
Together, our experiments and theoretical models indicate a distinct {regime} of \textit{swimming-limited} aggregation whose evolution is constrained by the intrinsic speed distribution of active agents, providing new insight into bacterial self-organization.

\end{abstract}

\maketitle

\renewcommand{\baselinestretch}{0.98}\normalsize

\newpage

\section{Introduction}

The dynamics of passive aggregation and fragmentation have been extensively studied due to their broad relevance to engineering, biology, and the physical sciences. Representative examples from the molecular to the celestial scale include polymerization \cite{MiddletonWiney2017}, colloidal aggregation \cite{LuWeitz2013}, aerosol dynamics \cite{Bourouiba2021}, fluid fragmentation \cite{Villermaux2007}, marine particle aggregation \cite{BurdJackson2009_marine}, as well as rainfall  and planet formation \cite{PumirWilkinson2016_cloud}.  
{In particular, colloidal aggregation is governed by the success rate of collisions, which distinguishes between the diffusion-limited regime}, in which every encounter leads to permanent bonding, and the reaction-limited {regime}, in which many pairwise encounters between clusters are required before they fuse into a larger cluster \cite{Weitz1984_DLA, FamilyMeakinVicsek1985,MeakinFamily1987,MeakinFamily1988}.

{In both passive and active systems, aggregation dynamics are strongly influenced by interactions between agents and their environment. For instance, turbulent flows drive the formation and fragmentation of fibre aggregates \cite{Verhille2025}, as well as the clustering and segregation of non-motile oceanic phytoplankton \cite{Borgnino2019}. Similarly, shear forces can restructure freshwater cyanobacterial colonies \cite{Sinzato_2025}. More complex environments, including nematic liquid crystals and active bacterial baths, can also induce the aggregation of passive colloidal particles into self-assembled microstructures \cite{Musevic2008_microstructures,Musevic2018_chains,Grober2023}. Bacteria alone display a remarkably diverse range of aggregation behaviours depending on their environment.}  In quiescent fluids, biofilm formation is typically driven by bridging and depletion attractions between bacteria \cite{Maier2021}, while under strong flow conditions bacteria can form biofilm streamers anchored to solid surfaces by extracellular polymeric substances (EPS) \cite{Savorana2025StressHardening}. Recent experiments further suggest that yeast and non-motile bacteria can generate self-amplifying flows that fragment and disperse cellular aggregates in a process termed circulation-driven aggregation \cite{David_Thutupalli_2025}.  Finally, in thin water layers, capillary forces drive the self-organization of non-motile bacteria into droplets and streamers \cite{Black2025Capillary}. {Although these examples mostly concern bacterial aggregation in isotropic Newtonian fluids,} many of the environments that bacteria naturally encounter, such as host mucus and the EPS secreted during biofilm formation, are anisotropic \cite{Figueroa-Morales2019,Repula2022}.

Liquid crystals (LCs) are widely used as model systems for anisotropic, structured fluids, in which the dynamics and growth of bacteria and undulatory  microswimmers have been extensively studied in recent years both experimentally and theoretically
\cite{Zhou2014_livingLC,Baza2026_livingLC,Mushenheim2014,Goral2022,Prabhune2024_tugofoars,Cheon2024,Lin2025_lyotropic,Quan2026,DattaSpagnoli2025_bacteriaLCs}. 
In particular, nematic LCs provide an exciting platform for studying active aggregation. In contrast to isotropic media where bacteria perform three-dimensional (3D) random walks \cite{BergBrown1972}, nematic LCs constrain bacterial motion to one dimension (1D), thereby increasing the likelihood of collisions between swimmers and their subsequent aggregation. 
Additionally, the elastic interactions mediated by the LC generate attractive forces that significantly reduce the likelihood of fragmentation (Fig.~\ref{fig:exp} a). Previous studies have demonstrated a strong mutual coupling between the collective dynamics of motile \textit{Bacillus subtilis} bacteria and the orientational order of LCs \cite{Zhou2014_livingLC,Baza2026_livingLC}, while \textit{Proteus mirabilis} were observed to form dynamic and reversible chain-like assemblies when placed in nematic LCs \cite{Mushenheim2014}. The orientational order of LCs also disrupts the run-and-tumble behaviour of multiflagellated bacteria, leading to distinct patterns of directional switching in 1D \cite{Goral2022,Prabhune2024_tugofoars}.
To further advance our understanding of active aggregation, we investigated the collective dynamics of motile \textit{Escherichia coli} bacteria in biocompatible LCs.

In this study, we focus on the quantitative description of \textit{E.~coli} bacterial chains that swim collectively along the nematic director, as previously reported for motile \textit{P.~mirabilis} cells by \citet{Mushenheim2014}. We find that the rate of aggregation is limited by the relative velocities between swimming bacteria, leading to an active  aggregation time scale that is much faster than the one required for passive bacteria to aggregate purely due to the elastic attraction induced by {distortions of the liquid crystal director}.  By analysing thousands of bacterial trajectories, we obtain the swimming speed distributions of individual bacteria and bacterial chains. Notably, the average velocity of the bacterial chains increases with the chain length, in contrast to our intuition based on a back-of-the-envelope calculation assuming random sorting of bacteria into chains and a Purcell-type force-balance model for the swimming kinematics of a bacterial chain. To elucidate the physical origins of the length--speed dependence of bacterial chains, we perform agent-based simulations of the 1D aggregation of bacteria swimming deterministically along a single lane. Our numerical simulations suggest that the positive correlation between chain length and speed is likely due to an active and time-dependent self-sorting process, in which faster bacteria are preferentially incorporated into longer chains. This characteristic feature of ``swimming-limited aggregation'' is consistent with a minimal aggregation model involving a bacterium and its two nearest neighbours, which shows that faster individuals encounter their neighbours sooner and are thus more likely to be incorporated into longer chains formed downstream in the aggregation process. Crucially, \blue{cell-to-cell variations} in swimming speed provide the heterogeneity required for the effective self-sorting of bacteria into chains.

The manuscript is organized as follows. In \S\ref{sec:exp}, we describe the experimental methodology and observations of swimming bacterial chains in nematic liquid crystals. In \S\ref{sec:theory}, we describe the theoretical approaches and results, starting with insights gained from a minimal model of aggregation between three bacteria, and followed by quantitative predictions from one-dimensional agent-based simulations. We conclude this section with a systematic investigation of the role of different model parameters, highlighting the rich temporal dynamics of a novel {regime} of ``swimming-limited aggregation''. In the final section, \S\ref{sec:comparison}, we compare the experimental and theoretical results and discuss the limitations and possible extensions of our study.

\section{Experimental Results}
\label{sec:exp}

\begin{figure*}
\centering
\includegraphics[width=\textwidth]{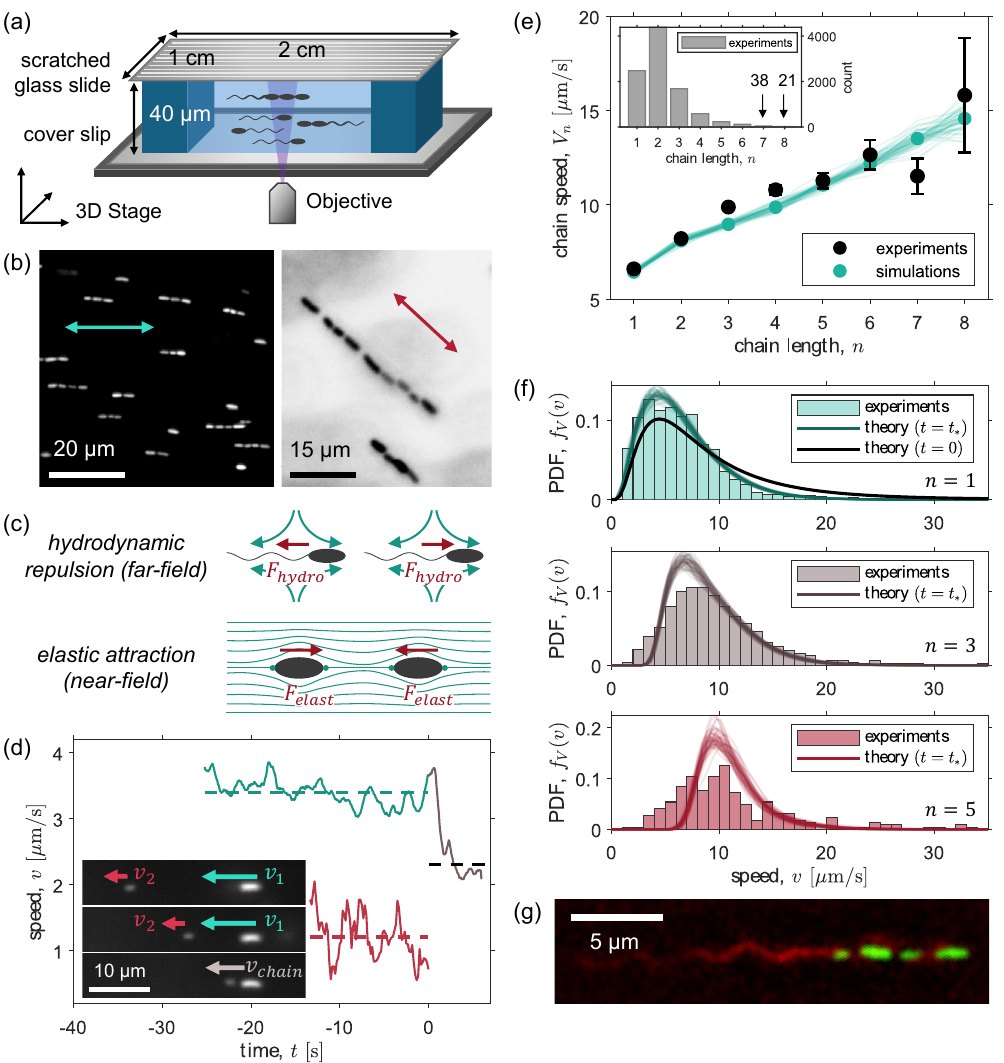}
\caption{(Colour online) \textbf{Aggregation dynamics of motile bacteria in nematic liquid crystals.} \\ 
(a) Illustration of the experimental setup showing \textit{Escherichia coli} (\textit{E.~coli}) bacteria swimming in an aligning chamber. The orientation of the nematic liquid crystal (LC) is imposed by nano-scratches on the glass slide.
(b) Micrographs of bacterial chains comprising two to nine bacteria that swim parallel to the nematic director, indicated by double arrows. (c) Competition between far-field hydrodynamic repulsion and near-field elastic attraction. (d) A newly merged bacterial chain swims near the average speed of the two bacteria prior to the encounter. Instantaneous (solid lines) and time-averaged (horizontal dashed lines) speeds of the two bacteria before encounter ($t=0$), followed by the measured (solid gray line) and predicted (dashed black line) speeds of the chain. (e) Average speed of bacterial chains against chain length, $n$, expressed as the number of bacteria. For the experiments, error bars indicate the standard error of the mean. For the theory, thin lines represent individual agent-based simulations evaluated at time $t_*$ (see Fig.~\ref{fig:simu}b), and circles indicate the average over 50 simulations. (Inset) Observation count of bacterial chains of length $n$. (f) Swimming speed distributions for bacterial chains of length $n=1, ~3, ~5$, based on sample sizes of $N_1= 2,470, ~N_3=1,682$ and $N_5 = 223$, respectively. Thin lines indicate theoretical predictions from independent agent-based simulations evaluated at time $t_*$,
while the black line indicates the initial lognormal distribution of bacterial speeds with mean $\overline{V}$ and variance $\sigma_{\text{\tiny V}}^2$. Simulation parameters for (e) and (f) are $\overline{V} ={\SI{9.18}{\micro\meter\per\second}}$ and $\sigma_{\text{\tiny V}} = {\SI{7.26}{\micro\meter\per\second}}$, with $\overline{V}$ estimated directly from the distribution of bacterial chain speeds and $\sigma_{\text{\tiny V}}$ obtained by a single-parameter fit to the average speed curve in (e). (g) ``Super-bundle'' configuration of flagella at the trailing edge of a chain of four bacteria with fluorescently labelled cell bodies (green) and flagella (red). \blue{Bacterial strains: \textit{E.~coli} {RP437} in panels (b), (d--f) and \textit{E. coli} {AD62} in panel (g).}}
\label{fig:exp}
\end{figure*}

We investigated the dynamics of motile \textit{E.~coli} bacteria immersed in a biocompatible nematic LC made of disodium cromoglycate (DSCG) dissolved in motility buffer. The mixture of bacteria and LC was introduced by capillarity into thin glass chambers of thickness $h = 40 \pm \SI{10}{\micro\meter}$, which were subsequently sealed to prevent evaporation-driven flows.
The orientation of the LC within the chamber was prescribed by nano-scratches on the upper glass surface, as shown in Fig.~\ref{fig:exp}(a). The nematic alignment constrained bacterial motion to a single preferred direction in the imaging plane, along 1D lanes parallel to the confining boundaries. Movies of fluorescent swimming bacteria were acquired on an inverted microscope fitted with a Hamamatsu Orca Flash 4.0 camera, typically using a 40× long working distance objective. Automated image analysis and tracking were then used to detect individual bacteria and bacterial chains, and to extract their trajectories, lengths and swimming speeds. The typical duration of the observations was around 5--\SI{7}{\minute}, and the observations were carried out {between \SI{2}{\minute} and \SI{30}{\minute}} after the bacteria were mixed in the nematic LC phase, corresponding to the onset of aggregation. Further details on the materials and methods used in experiments are provided in Appendix \ref{app:methods}.

We observed that swimming \textit{E.~coli} self-organize into chain-like assemblies that continue to swim collectively along the nematic director, as shown in Fig.~\ref{fig:exp}(b). The cell bodies are arranged in a closely packed, single-file configuration containing between two and nine bacteria, with rare instances of chains up to 16 bacteria long (not shown). This suggests that aggregation continues beyond the initial merging of bacterial pairs through successive encounters with other bacteria or pre-existing chains. The micrographs in Fig.~\ref{fig:exp}(b) provide direct qualitative evidence for the emergence of extended, dynamically assembled bacterial chains in the nematic LC. However, in the quantitative analysis that follows, we focus on bacterial chains of length up to eight, for which the sample size is sufficiently large. 

Bacteria encounter each other due to the broad \blue{cell-to-cell variations} in swimming speed, so that faster bacteria can catch up with slower ones in the longitudinal direction. The LC promotes these collisions by strongly confining the bacterial motion to a one-dimensional path, which prevents swimmers from overtaking each other laterally, \blue{unless they are moving in separate ``lanes'' (see Fig.~\ref{figS:sliding})}. Due to their characteristic pusher-type flows, swimming bacteria generate far-field hydrodynamic interaction forces that are repulsive along the swimming axis and, thus, the nematic director (see Fig.~\ref{fig:exp}(c), upper panel). At sufficiently small separations, however, the dynamics become dominated by near-field elastic interactions arising from local distortions of the director field. Each bacterium distorts the local alignment of the LC mainly through its cell body, which is much wider than the flagellar filaments (Fig.~\ref{fig:exp}(c), lower panel). In particular, spheroidal inclusions with planar anchoring conditions induce a bipolar distortion of the nematic LC with two surface topological defects termed ``boojums'' at opposite poles of the particle \cite{VolovikLavrentovich1983_boojums,Terentjev1995_boojums}. These defects  are represented as dots in Fig.~\ref{fig:exp}(c). The overlap of these distortion fields reduces the total elastic free energy stored in the nematic LC when the cells come together, {thereby inducing} an effective attraction between cells that favours aggregation and stabilizes the newly formed bacterial chain.

Passive particles also experience attractive forces in nematic LCs through the same type of LC distortions described above, leading to their self-assembly into linear chains \cite{Musevic2018_chains}. Therefore, a natural question arises: {do active bacteria form dimers faster than passive bacteria?} We can distinguish the active aggregation of motile bacteria from the passive aggregation of colloids in nematic LCs by considering the balance between elastic forces and viscous drag on the cell bodies of non-motile bacteria.
\blue{We estimate that, under the low bacterial density conditions of our experiments, non-motile bacteria would passively aggregate through LC-mediated attractive interactions only over time scales of hours or longer (see details in Appendix \ref{app:estimation_passive})}. {By contrast, active swimmers are expected to meet within minutes. For bacteria with a standard deviation in speed of $\sigma_{\text{\tiny V}} = \SI{7}{\micro\meter\per\second}$, later shown to be consistent with our experiments, and a typical initial separation of $d = \SI{1}{\milli\meter}$, the expected encounter time is approximately $d/\sigma_{\text{\tiny V}} \approx \SI{2.4}{\minute}$. The separation of $\SI{1}{\milli\meter}$  corresponds to the bacterial area density in our samples, $c^{2D}_{bact} = 10^3~\SI{ }{\per\milli\meter\squared}$, if we assume that bacteria swim along effective $\SI{1}{\micro\meter}$-wide ``lanes'' set by the width of the cell body (see Fig.~\ref{figS:sliding}). The estimated time scale for active aggregation is thus consistent} with our observation of long bacterial chains formed between \SI{2}{\minute} and \SI{30}{\minute} from the onset of aggregation, suggesting that the intrinsic activity of bacteria plays a key role in {this process}.

Direct observations of merging events are not readily available since encounters between bacteria are rare, and only exceptionally captured in the observation window. Nevertheless, Fig.~\ref{fig:exp}(d) shows a merging event in which two bacteria meet and continue swimming together as a coherent chain due to \blue{physical constraints from the nematic LC}. This newly formed chain swims at a speed close to the average of the individual bacterial speeds before the encounter.
The ``rear-end'' collision shown here, in which a faster bacterium catches up with a slower one, complements the earlier observation of a ``head-on'' collision between two \textit{P.~mirabilis} cells {swimming in opposite directions}, reported by \citet{Mushenheim2014}. As we demonstrate later in the manuscript, these two observations of bacterial encounters support a ``dry mechanical coupling'' model of bacterial propulsion in nematic LCs, detailed in Appendix~\ref{app:drymech}. In this model, the total flagellar thrust and viscous drag of the emerging chain are the sums of the individual contributions of the precursor bacteria. Hence, the theoretically predicted chain speed is equal to the average of their individual speeds, consistent with the measurements in Fig.~\ref{fig:exp}(d). 

We next examine the swimming behaviour of bacterial chains of length $n$, measured as the number of bacteria within the chain rather than its physical length. As shown in Fig.~\ref{fig:exp}(e), we observe a significant increase in the average swimming speed with increasing chain length. 
Note that the mean chain length in our experimental dataset is $\langle n\rangle \approx 2.2$ bacteria (Fig.~\ref{fig:exp}(e), inset), which we later use as a reference point to estimate the age of the aggregation process. 

At first sight, the positive correlation between chain length and speed is surprising, since the pairwise merging event discussed earlier suggests that a newly formed chain should swim at approximately the average speed of its constituent parts. However, for each chain length, the swimming speeds remain broadly distributed around the mean, as shown in Fig.~\ref{fig:exp}(f). 
The unimodal, right-skewed shape of these distributions suggests that, among the fundamental continuous probability distributions, the lognormal distribution is well-suited to describe the speeds of individual bacteria. We will therefore use this distribution later in the theoretical modelling section.
Typical values of mean speed and standard deviation for solitary swimming bacteria were approximately $\SI{7}{\micro\meter\per\second}$ and $ \SI{5}{\micro\meter\per\second}$, respectively (Fig.~\ref{fig:exp}(f), top panel).
Overall, Fig.~\ref{fig:exp}(f) illustrates that the \blue{cell-to-cell variation in swimming speed} is inherited by longer chains. This suggests that the same variability that enables bacteria to encounter each other may also determine how they self-organize into chains. In the following section, we rationalize this mechanism of ``swimming-limited aggregation'' using theoretical models and consider the possibility that bacteria are not incorporated into chains at random but undergo a dynamic self-sorting process based on their speeds.

Finally, we briefly report additional experimental observations obtained using a dual fluorescent strain of \textit{E.~coli}, which allows the simultaneous visualization of bacterial cell bodies and flagella. In these experiments, bacterial chains were seen to feature a group of closely spaced cell bodies at the leading edge and distinctive configurations of helical flagellar filaments from multiple bacteria, as shown in Fig.~\ref{fig:exp}(g) and in further examples in Supp.~Fig.~\ref{figS:super-bundle}. In the example of Fig.~\ref{fig:exp}(g), the flagella are organized in a coherent ``super-bundle'' trailing behind the chain, while the closely packed cell bodies remain aligned with the nematic director. The continued rotation of this trailing flagellar bundle provides a plausible propulsion mechanism for the entire chain, allowing it to swim collectively as a single unit. Hence, these images provide direct physical insight into how bacterial chains remain motile after aggregation and offer further support to the theoretical model introduced in the next section.

\section{Theoretical Modelling}
\label{sec:theory}

To elucidate the physical mechanism behind this active process of bacterial aggregation, we propose two theoretical model systems. The first considers the probability of encounter between a bacterium and its two nearest neighbours along the swimming direction, i.e., the nematic director. The second model extends these ideas to many bacteria with variable swimming speeds that aggregate deterministically along a single ``swimming lane'' in the nematic LC. This model connects the encounter probabilities at the nearest-neighbour level to the dynamic self-organization of bacteria into chains at the macroscopic level. 

\subsection{Minimal Model of Aggregation}

\begin{figure}
\centering
\includegraphics[width=.5\textwidth]{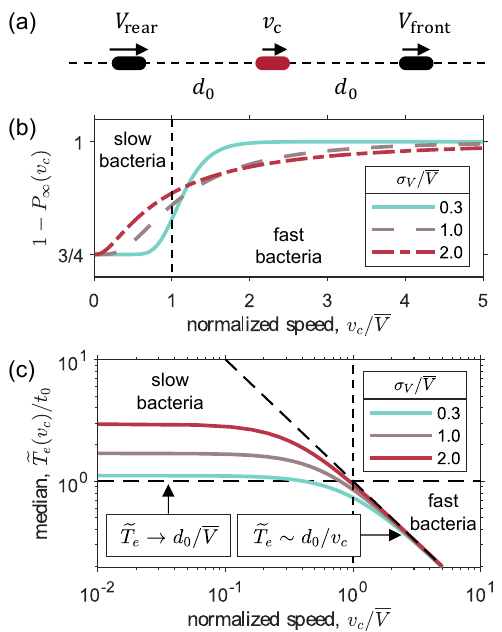}
\caption{(Colour online) \textbf{Encounter probabilities in a minimal model of bacterial aggregation.} 
\\ (a) Schematic of the model. The central bacterium moves at a given speed, $v_c>0$, while the swimming speeds of the rear and front neighbours are random variables sampled from a lognormal distribution with mean $\overline{V} = \SI{9}{\micro\meter\per\second}$ and $\sigma_{\text{\tiny V}}$ indicated in the legend. The peripheral bacteria have equal probabilities of moving forward or backward. 
(b) Probability of finite-time encounter, and (c) median encounter time as a function of $v_c$. {Vertical lines separate the limiting behaviours of slow ($v_c < \overline{V}$) and fast bacteria.} Times are normalized by the characteristic time scale of aggregation, $t_0 = d_0/\overline{V}$, where $d_0$ is the initial distance between bacteria.}
\label{fig:prob}    
\end{figure}

We begin by considering a bacterium and its two nearest neighbours along the swimming direction, Fig.~\ref{fig:prob}(a). The central bacterium swims forward with a given velocity, $v_c \geq 0$, while the speeds of the rear and front neighbour, $|V_{\mathrm{rear}}|$ and $|V_{\mathrm{front}}|$, are independent and identically distributed random variables sampled from a lognormal distribution with mean $\overline{V}$ and standard deviation $\sigma_{\text{\tiny V}}$, consistent with experimental observations. The peripheral bacteria are equally likely to swim forward or backward relative to the swimming direction of the central bacterium, so that  $\sgn(V_{\mathrm{rear}})$ and $\sgn(V_{\mathrm{front}})$ take the values $\pm1$ with equal probability. How long do we expect to wait for an encounter between two bacteria in this minimal model of aggregation?

The main assumption of our model is that the leading-order kinematics are governed by the relative velocities between bacteria, which are themselves determined by the normalized speed of the central bacterium, $v_c/\overline{V}$, and the speed variability of its nearest neighbours, $\sigma_{\text{\tiny V}}/\overline{V}$. The model neglects hydrodynamic interactions or attractive forces due to the LC, as well as fluctuations in the bacterial speeds. As detailed in Appendix \ref{app:probab}, we find that the probability of waiting longer than a time $t$ to observe an encounter between the central bacterium and one of its nearest neighbours, conditional on the central bacterium having speed $v_c$, is given by
\begin{equation}
    p(T_e \geq t) = \left[\frac{1}{2}+\frac{1}{2}F_V\left(v_c + \frac{d_0}{t}\right)\right]\left[\frac{1}{2}-\frac{1}{2}\text{sgn}\left(v_c-\frac{d_0}{t}\right)F_V\left(\left|v_c-\frac{d_0}{t}\right|\right)\right],
    \label{eq:Te}
\end{equation}
where the random variable $T_e$ represents the first encounter time, $d_0$ is the initial distance between bacteria, and $F_V(v) = p(V\leq v)$ is the cumulative distribution function of individual bacterial speeds. The first and second expressions in square brackets on the right-hand side of Eq.~\eqref{eq:Te} correspond to the probabilities that the central bacterium takes longer than time $t$ to encounter its rear and front neighbours, respectively. {In the first set of brackets, the first term represents the probability that the rear neighbour swims away from the central bacterium, while the second term denotes the probability that it swims towards the central bacterium, but not fast enough to catch up with it within time $t$. The second set of brackets denotes the probability that the front neighbour swims away from the central bacterium fast enough to avoid being caught within time $t$.}

As required, the analytical expression for  $p(T_e \geq t)$ decreases monotonically in time between the limits of 1 as $t\to 0$ and $\frac{1}{4}(1-F_V(v_c)^2)$ as $t\to\infty$, since the cumulative distribution function $F_V(v)$ is a monotonically increasing function of $v$. Hence, there is a finite probability that \blue{the central bacterium} will never meet \blue{its neighbours}, given by
\begin{equation}
    P_\infty(v_c) = \lim_{t\to\infty} p(T_e \geq t) = \frac{1}{4}\left(1-F_V (v_c)^2\right),
    \label{eq:Pinf}
\end{equation}
where $F_V (v_c)$ is the probability that a bacterium has speed at most $v_c$.
The probability quantified in Eq.~\eqref{eq:Pinf} corresponds to the joint scenario in which the front neighbour moves forward faster than the central bacterium, i.e., $V_{\mathrm{front}}\geq v_c \geq 0$ with probability $\frac{1}{2}\left[1-F_V (v_c)\right]$, and the rear bacterium either moves backward or moves forward more slowly than the central bacterium, i.e., $V_{\mathrm{rear}}<0$ or $0\leq V_{\mathrm{rear}} \leq v_c$ with total probability $\frac{1}{2}\left[1+F_V (v_c)\right]$. The product of these two probabilities gives the right-hand side of Eq.~\eqref{eq:Pinf}. Since $F_V(v_c)$ is a monotonically increasing function of $v_c$, we deduce that the probability of infinite encounter time, $P_\infty(v_c)$, decreases with the speed of the central bacterium.

Our minimal model predicts that faster bacteria are always more likely to meet one of their neighbours in finite time, as shown in Fig.~\ref{fig:prob}(b). The curves increase monotonically between the limiting values of $p(T_e<\infty)\to 3/4$ as $v_c\to 0$ and $p(T_e<\infty)\to1$ as $v_c\to \infty$, with the sigmoidal shape becoming steeper as the standard deviation of the speed distribution decreases. 
{The lower limit of $3/4$ reflects the fact that only one of the four equally probable combinations of swimming directions prevents a future encounter.}
Here, the speeds of neighbouring bacteria are drawn from a lognormal distribution with mean $\overline{V} = \SI{9}{\micro\meter\per\second}$ and standard deviation $\sigma_{\text{\tiny V}} \in \{2.7, 9, 18\}~\SI{}{\micro\meter\per\second}$, and the finite-time encounter probability is computed numerically using Eq.~\eqref{eq:Pinf}. A smaller standard deviation of bacterial speeds corresponds to a sharper transition from zero to one in the cumulative distribution function, $F_V$, resulting in a steeper sigmoidal curve for the finite-time encounter probability.

Since there is a finite probability that the three bacteria will never meet, the expected value of the first encounter time is undefined. However, a useful metric to consider in such cases is the median encounter time, $\widetilde{T}_e$, for which $p(T_e\leq\widetilde{T}_e)=1/2$. The median is well defined because $p(T_e<\infty) = \frac{1}{4}\left(3+F_V (v_c)^2\right)>1/2$. The observation that the median encounter time decreases monotonically with the speed of the central bacterium, as shown in Fig.~\ref{fig:prob}(c), reinforces the notion that faster bacteria merge with other bacteria and form swimming chains more rapidly than slower bacteria. Furthermore, the median first encounter time increases with the standard deviation of the bacterial speed distribution, a trend that may be related to the waiting-time paradox \cite{MasudaHiraoka2020}. {This effect, also known as the bus paradox, states that the expected waiting time of a passenger who arrives at a bus stop without inspecting the timetable is at least half the mean interval between consecutive buses. The lower bound is attained only when buses arrive at perfectly regular intervals; otherwise, the mean waiting time increases with the variance of time intervals between consecutive buses. Similarly, from the perspective of the central bacterium, the waiting time before an encounter depends on the relative velocity of its neighbours and is therefore expected to increase with the standard deviation of the speed distribution from which those velocities are sampled.}

In addition to the properties noted above, the median encounter time exhibits two asymptotic behaviours, highlighted by the boxes in Fig.~\ref{fig:prob}(c). Exceptionally fast bacteria ($v_c\gg \overline{V}$) encounter their front neighbour on a time scale approximately equal to the initial separation divided by their intrinsic speed. Meanwhile, exceptionally slow bacteria ($v_c \ll \overline{V}$) must wait for one of their neighbours to come to them. For very narrow speed distributions, this waiting time is simply the initial separation divided by the mean bacterial speed.

In the agent-based simulations that follow, we extend this framework to include the swimming dynamics of many bacteria swimming along a single ``lane'' within the LC. In this setting, an encounter between a reference bacterium and its rear neighbour may be accelerated if the rear neighbour itself is reached from behind by a faster bacterium. Such a precursor merging event would increase the swimming speed of the agent immediately behind the reference bacterium, thereby reducing the expected encounter time between it and its rear neighbour. A similar argument applies to interactions with the front neighbour, which may encounter a slower bacterium swimming further ahead and slow down as a result of the merging event. Hence, the encounter times predicted by the minimal model of aggregation for an isolated trio of bacteria provide an upper bound on the first encounter time in many-body systems.

\subsection{Agent-Based Simulations}
\label{sec:ABS}

{Having established that encounter times decrease with both increasing swimming speed and decreasing speed variability}, we now examine the dynamics emerging from these biased encounter probabilities. How do bacteria behave after merging into a chain, and how does aggregation proceed beyond this stage? 
Using a modelling approach similar to that of \citet{Purcell1977}, we consider a ``dry'' mechanical coupling model in which the swimming dynamics of a bacterial chain are governed by the balance of forces along the direction of swimming, detailed in Appendix \ref{app:drymech}. The model is called ``dry'' as it does not account for the hydrodynamic interactions between bacteria or the elastic forces due to the LC. 
We further assume that flagellar bundles rearrange instantaneously at the trailing edge of a bacterial chain during head-on collisions, and that all bacteria have the same net drag coefficient. The latter assumption enables us to use the bacterial speed distributions measured in experiments, Fig.~\ref{fig:exp}(f), without knowing the underlying distributions of thrust and drag coefficients among bacteria. 

By balancing the total thrust generated by flagellar filaments with the viscous drag on the cell bodies and the flagella, our model predicts that a chain of bacteria with individual swimming speeds $v_1, \dots, v_n$ should move at the average speed of its constituent bacteria,
\begin{equation}
    v_{\text{chain}} = \frac{1}{n}\sum_{k=1}^n v_k.
    \label{eq:vchain}
\end{equation}
The underlying assumptions and complete derivation of this result are detailed in Appendix \ref{app:drymech}.
In particular, when a chain of length $n_1$ and speed $\tilde{v}_1$  merges with a chain of length $n_2$ and speed $\tilde{v}_2$, they form a chain of length $n=n_1 + n_2$ with swimming speed
\begin{equation}
    v_{\text{chain}} = \frac{n_1\tilde{v}_1+n_2\tilde{v}_2}{n},
    \label{eq:vchain_n1n2}
\end{equation}
where the speeds of the precursor chains can be related to their constituent parts as per Eq.~\eqref{eq:vchain}. 

The new direction of swimming after a merging event is trivial in the case of a rear-end collision, where both agents swim in the same direction before the encounter, see Fig.~\ref{fig:simu}(a). For head-on collisions, the direction is set by the strongest participant in the encounter, such that the signed velocities of the precursor and resulting chains are related by 
\begin{equation}
    V_{\text{chain}} = \mathrm{sgn}(n_1\widetilde{V}_1+n_2\widetilde{V}_2)\frac{n_1|\widetilde{V}_1|+n_2|\widetilde{V}_2|}{n}.
    \label{eq:Vchain}
\end{equation}
Eq.~\eqref{eq:Vchain} constitutes the kinematic ``rule of engagement'' underlying our agent-based simulations.  

{In the simulations, we capture the \blue{cell-to-cell variations} in bacterial swimming speeds by sampling the intrinsic speeds of active agents from a lognormal distribution with parameters $\mu$ and $\sigma$. These parameters are uniquely related to the mean, $\overline{V}$, and standard deviation, $\sigma_{\text{\tiny V}}$, of the distribution through $\overline{V} = \exp\left(\mu+\sigma^2/2\right)$ and $\sigma_{\text{\tiny V}}/\overline{V} = \sqrt{\exp(\sigma^2)-1}$. As previously mentioned, the unimodal, right-skewed shape of the lognormal distribution makes it well-suited for describing the speed distributions measured experimentally, Fig.~\ref{fig:exp}(f). Using a parametric distribution, rather than an empirical model fitted directly to the experimental data, allows us to perform a systematic parametric study and disentangle the effects of different physical parameters. In particular, the standard deviation of the distribution captures the variability in swimming speeds, which, as shown below, plays a crucial role in this active aggregation process.}

The setup of our agent-based simulations of bacterial aggregation is depicted in Fig.~\ref{fig:simu}(a). We initialize 10,000 bacteria as equally spaced point particles along a single ``swimming lane'' in the nematic LC. Each bacterium has an equal probability of swimming left or right, and their speeds are sampled from a lognormal distribution with mean $\overline{V}$ and standard deviation $\sigma_{\text{\tiny V}}$ {consistent with typical experimental values. We use $\overline{V} = \SI{9}{\micro\meter\per\second}$ and $\sigma_{\text{\tiny V}} = \SI{6}{\micro\meter\per\second}$ unless stated otherwise}. Active agents maintain a constant swimming velocity until they encounter another agent, when they instantaneously form a longer chain with a velocity given by the dry mechanical coupling model in Eq.~\eqref{eq:Vchain}. Since the model treats all swimming agents as point particles and all mergers as instantaneous events, the qualitative features of aggregation do not depend on the average initial separation between bacteria, $d_0$. The initial separation simply rescales the distance travelled and the time elapsed between successive encounters, {meaning that the aggregation process develops on a} characteristic time scale given by $t_0 = d_0/\overline{V}$.

\begin{figure*}
\centering
\includegraphics[width=\textwidth]{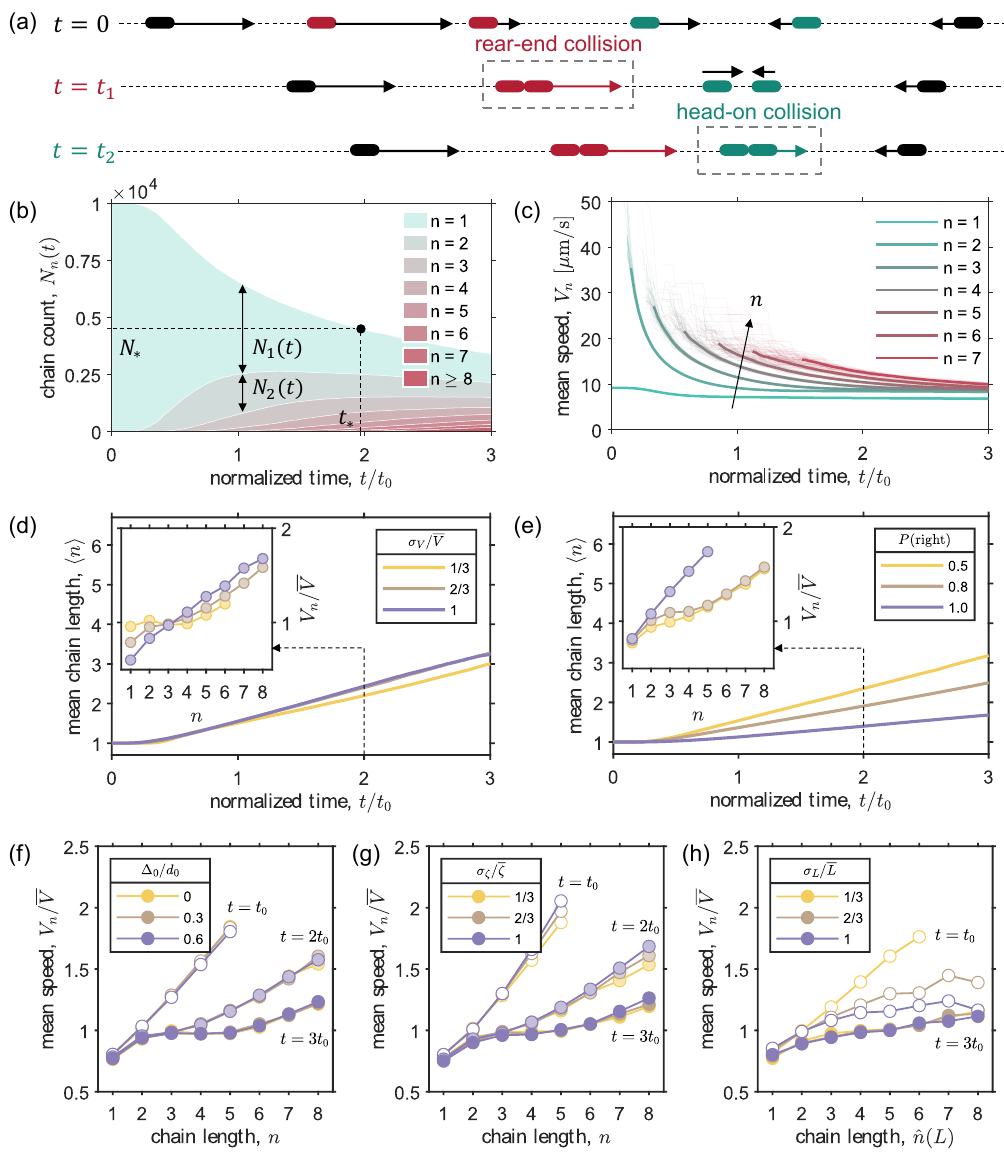}
\caption{(Colour online) \textbf{Theoretical aggregation dynamics in agent-based numerical simulations.} (a) Schematic of single-lane swimming simulations showing rear-end and head-on collisions at times $t_1$ and $t_2$, respectively. Unless otherwise stated in the caption, the initial speeds of $10^4$ bacteria were sampled from a lognormal distribution with mean $\overline{V} = \SI{9}{\micro\meter\per\second}$ and standard deviation $\sigma_{\text{\tiny V}} = \SI{6}{\micro\meter\per\second}$, {consistent with experiments}, and their starting positions were evenly spaced at separation $d_0$. Times are normalized by the characteristic time scale of aggregation, $t_0 = d_0/\overline{V}$. 
(b) Stacked band plot showing the number of $n$-length chains, $N_n(t)$, during a single simulation. Time $t_*$ indicates the moment when the \blue{total chain count} reaches 45\% of the initial number of bacteria (i.e., $N_* = 4.5\times 10^3$). This corresponds to a mean chain length $\langle n\rangle \approx 2.2$ bacteria, as in the experiments.
(c) Time evolution of the average speed of $n$-length chains, $V_n(t)$. Thin lines represent independent simulations, and thick lines show averages over 50 simulations, {starting once $n$-length chains have appeared in all simulations.}
Across panels (d-h), the results are averaged over ten independent simulations for each set of parameters.  
{(d) Effect of speed variability, quantified through the standard deviation, $\sigma_{\text{\tiny V}}$, of the bacterial speed distribution, on the time evolution of the mean chain length. (Inset) Corresponding speed-length dependence of bacterial chains at $t=2t_0$.} Bacterial speeds were sampled from a lognormal distribution with mean $\overline{V} = \SI{9}{\micro\meter\per\second}$ and standard deviation $\sigma_{\text{\tiny V}}$ indicated in the legend. 
{(e) Effect of directional bias, quantified through the right-swimming probability $P(\text{right})$, on the time evolution of the mean chain length. (Inset) Corresponding speed-length dependence of bacterial chains at $t=2t_0$.} 
(continued on next page)
}
\label{fig:simu}
\end{figure*}
\addtocounter{figure}{-1}
\begin{figure*}
\caption{(continued) \textbf{Theoretical aggregation dynamics in agent-based numerical simulations.}
(f)  Effect of irregular spatial distribution on the speed--length dependence of bacterial chains. Initial bacterial positions were displaced from a regular grid of spacing $d_0$ by a random amount sampled from a uniform distribution on the interval $[-\Delta_0/2, \Delta_0/2]$, with $\Delta_0$ indicated in the legend.
(g) Effect of drag-coefficient distribution on the speed--length dependence of bacterial chains. Individual bacterial drag coefficients were drawn from a lognormal distribution with rescaled mean $\overline{\zeta}=1$ and standard deviation $\sigma_\zeta$ indicated in the legend. 
(h) Effect of cell-length distribution on the speed--length dependence of bacterial chains, through its influence on both the drag coefficient, via $\zeta(L) \propto L$, and the apparent chain length, $\hat{n}(L)$, taken to be the nearest positive integer to $L/\overline{L}$. Individual cell lengths were drawn from a lognormal distribution with rescaled mean $\overline{L}=1$ and standard deviation $\sigma_L$ indicated in the legend.
}
\end{figure*}

The simulations provide insight into the rich temporal dynamics of the numbers and speeds of bacterial chains, as shown in Fig.~\ref{fig:simu}(b--c). First, we note that bacteria aggregate into longer and longer chains, Fig.~\ref{fig:simu}(b), with the remaining number of active agents {(including both chains and solitary bacteria)} falling under half of the original number of bacteria by time $2t_0$. In the simulations, the number of solitary swimming bacteria, $N_1$, decreases monotonically over time, whereas the number of dimers, $N_2$, increases initially and then decreases, when the rate of bacterial pairs merging with other agents exceeds the rate of solitary bacteria merging into pairs. Similar dynamics were observed for longer chains. The stacked band plot illustrates the macroscopic transport of bacteria between a discrete number of communicating vessels or compartments of $n$-length chains. In the absence of fragmentation, the fluxes between compartments are irreversible and no steady state is reached. 

Since bacteria self-organize into chains based on their relative velocities and probabilities of encounter, this is reflected in the time evolution of the mean speed of $n$-length chains, Fig.~\ref{fig:simu}(c). We observe that the average speed in each compartment decreases monotonically in time, and that longer chains always swim faster on average, $V_n(t) > V_m(t)$ for any $n>m$. To interpret this observation, note that the speed of our simulated bacterial chains is merely the average of the speeds of their constituent parts, Eq.~\eqref{eq:vchain}. Hence, $V_n(t)$ is a measure of the average speed of all bacteria in $n$-length chains at time $t$, suggesting that faster bacteria are preferentially found in longer chains. This is consistent with our minimal model of aggregation which predicts that the {median} time an agent encounters one of its neighbours is a decreasing function of its speed, Fig.~\ref{fig:prob}(c). By extension, we expect that the rate at which active agents transition into longer-chain-length compartments should increase with the speed of the active agent. Hence, the average speed within any given compartment decreases over time because (i) the fastest agents continue to merge into even longer chains, and (ii) later arrivals into the compartment are composed of slower bacteria than those incorporated during earlier merging events.

Our agent-based simulations reveal a dynamic, self-organized aggregation process in which both the number and speed of bacterial chains evolve continuously in time (Fig.~\ref{fig:simu}(b--c)). This is also reflected in the sustained growth of the average chain length over time (Fig.~\ref{fig:simu}(d--e), main panels), confirming that the system does not approach a steady state, as expected for a process driven by irreversible aggregation. At the same time, {the average speed of longer chains is consistently faster than that of shorter ones} (Fig.~\ref{fig:simu}(d--e), inset panels), indicating that the active agents are dynamically sorted into chains according to their swimming speeds. Notably, the temporal reduction of average chain speeds described in Fig.~\ref{fig:simu}(c) implies that the speed--length relation must be age dependent, and its slope depends on the time when the system is observed, as shown in Fig.~\ref{fig:simu}(f--h).

In  Fig.~\ref{fig:simu}(d--e), we \blue{also} demonstrate that the aggregation process is fundamentally shaped by the velocity distribution of active agents, which we control through the standard deviation of their intrinsic speeds, $\sigma_{\text{\tiny V}}$, and the initial probability of swimming towards the right, $P(\text{right})$. 

First, we observe that speed variability has only a weak effect on the growth of the average chain length (Fig.~\ref{fig:simu}(d), main) but a significant impact on the speed--length relation (Fig.~\ref{fig:simu}(d), inset). {Crucially}, populations with larger $\sigma_{\text{\tiny V}}$ exhibit a stronger positive correlation between chain length, $n$, and normalized mean speed, $V_n/\overline{V}$, consistent with the fact that a larger speed variability allows faster bacteria to aggregate more easily and thus favours their incorporation into longer chains. 
For sufficiently narrow speed distributions, {the self-sorting mechanism} becomes weaker and may even produce non-monotonic speed--length relations, with a local minimum near $n=3$ observed in simulations when $\sigma_{\text{\tiny V}}/\overline{V} = 1/3$. Additional simulations (not shown) suggest that this feature propagates towards larger $n$ as the simulation time increases. {The observed non-monotonicity can be understood as a competition between dynamic speed sorting and compositional averaging. For $\sigma_{\text{\tiny V}}=1/3$, all particles swim at similar speeds. This weakens the sorting mechanism, as fast swimmers are only marginally more likely to aggregate than slower ones. In this regime, the speed of a chain becomes sensitive to stochastic fluctuations in composition. Short chains may be slightly enriched in faster swimmers, while intermediate-length chains tend to average out these speed differences. {The few long chains that emerge under these conditions are necessarily enriched in exceptionally fast swimmers}, which may explain the non-monotonic speed--length relation at $\sigma_{\text{\tiny V}}=1/3$.}

{Next, we find that} the directional bias of swimmers, quantified by the right-swimming probability $P(\text{right})$, strongly influences both the average chain length (Fig.~\ref{fig:simu}(e), main) and the speed--length relation (Fig.~\ref{fig:simu}(e), inset). Notably, chains take longer to form when bacteria swim predominantly in one direction, since fewer head-on collisions take place. However, the chains that do emerge within a given time window (here, $t = 2t_0$) exhibit a stronger speed--length correlation than in an unbiased system. This is because, in a strongly biased system, chains form only when exceptionally fast swimmers catch slower ones from behind, whereas in a symmetric system, cells swimming in opposite directions may aggregate and form chains even if their intrinsic speeds are not exceptionally fast. 

Finally, in Fig.~\ref{fig:simu}(f--h), we test the robustness of the speed--length relation to changes in the initial arrangement of bacteria and the variability of the bacterial drag coefficient, $\zeta$, and cell length, $L$. As shown in Fig.~\ref{fig:simu}(f), random displacements of the initial bacterial positions about a fixed mean separation have a negligible effect on the speed--length relation. A similarly weak influence is observed for variations in drag coefficient. In Fig.~\ref{fig:simu}(g), we relax the assumption of identical drag coefficients and model the chain speed as $v_{\text{chain}} = \sum_{k=1}^n \zeta_kv_k/\sum_{k=1}^n \zeta_k$ (cf. Eq.~\eqref{eq:vchain}), where the  individual speeds and drag coefficients, $v_k$ and $\zeta_k$, are drawn from lognormal distributions with means $\overline{V}$ and $\overline{\zeta}$ and standard deviations $\sigma_{\text{\tiny V}}$ and $\sigma_\zeta$, respectively. Notably, variations in drag coefficient have a much weaker effect on the speed--length curve than comparable variations in swimming speed (compare Fig.~\ref{fig:simu}(d) and (g) at $t=2t_0$). In Fig.~\ref{fig:simu}(h), we further consider the possibility that drag heterogeneity arises from variations in cell length, so that \blue{cell-to-cell variations} affect both the swimming kinematics and the mapping between physical chain length and the apparent bacterium count, $\hat{n}$. Specifically, we sample cell lengths from a lognormal distribution with mean $\overline{L}$ and standard deviation $\sigma_{\text{\tiny L}}$ and let the drag coefficient be proportional to cell length, $\zeta(L) = \zeta_0 L$, rescaling $\zeta_0 = 1$ for convenience. We also define the apparent chain length as the nearest positive integer to $L/\overline{L}$, which is relevant to the automated image-analysis procedure used in experiments. The results in Fig.~\ref{fig:simu}(h) suggest that cell-length variability can influence the apparent speed--length relation, especially at earlier times in the simulation. Comparison of panels (g) and (h) indicates that this change is due primarily to the automated segmentation of chains into bacteria of average length $\overline{L}$, rather than from the corresponding variations in drag coefficient. 

Before closing this section, we comment further on the time scales and dynamical features of this active aggregation process. A common feature of Figs.~\ref{fig:simu}(f--h) is that the speed--length relation becomes shallower over time, which is consistent with the ageing effect already visible in Fig.~\ref{fig:simu}(c). Since the characteristic time scale of aggregation is set by $t_0=d_0/\overline{V}$, varying either the mean separation between bacteria or their mean speed would simply rescale the time between encounters, which justifies keeping these parameters constant in the simulations. Meanwhile, varying the observation time relative to $t_0$ changes the ``dynamical age'' of the aggregating system relative to the underlying rate of collisions, \blue{and therefore modifies the speed--length dependence predicted by the theory. We will revisit this point in the next section when we compare the results from experiments and simulations.}

Overall, this systematic parametric study shows that the aggregation dynamics observed theoretically are governed primarily by the velocity distribution of active agents, including both speed variability and directional bias, rather than by heterogeneity in drag coefficient, cell length, or initial spatial arrangement. This supports the conclusion that the positive correlation between chain speed and length in our numerical simulations emerges from a process of swimming-limited aggregation driven by \blue{cell-to-cell variations} in swimming speeds.

\section{Comparison of Experimental and Theoretical Results}
\label{sec:comparison}

In the previous two sections, we presented experimental observations of the aggregation and collective swimming of \textit{E.~coli} bacteria in nematic LCs, together with theoretical predictions from 1D agent-based simulations of active particles with intrinsic speed variability. We now discuss these results and compare experiment and theory quantitatively, focusing on the average speed--length dependence of bacterial chains. 

\blue{As shown in the previous section, the speed--length relation predicted by the theory depends on the ``dynamical age'' of the system. To enable a direct comparison between experiments and simulations, we must therefore introduce an objective measure of the age of the aggregating system. To this end, we define the mean chain length $\langle n \rangle = \sum_{k=1}^N n_k / N$, where $N$ is the total number of tracked objects (including both chains and solitary bacteria), and $n_k$ is the number of bacteria within each object. For the experiments, this average is calculated over the entire dataset, yielding $\langle n \rangle \approx 2.2$ bacteria per chain. Meanwhile, the simulations enable us to monitor the mean chain length over time. The critical value of $\langle n \rangle \approx 2.2$ is reached at time $t_*$, indicated in Fig.~\ref{fig:simu}(b), when the total chain count, $N^*$, reaches approximately 45\% of the initial number of swimmers in the simulations. 
This is the appropriate simulation time point at which theoretical predictions and experimental observations can be compared.}

We also established, \blue{in the previous section}, that the initial spatial arrangement and heterogeneity in drag coefficient play only a secondary role in the theoretical aggregation dynamics. This leaves the initial speed distribution of swimmers as the key set of parameters to determine. Since no left--right symmetry breaking is observed in the swimming dynamics of \textit{E.~coli} in DSCG solution, we assume $P(\text{right})=0.5$ in the simulations. By inspecting the distributions in Fig.~\ref{fig:exp}(f), we find that a lognormal distribution provides a reasonable model for the intrinsic speeds of active particles initialized in our numerical simulations, leading to two fitting parameters: the mean, $\overline{V}$, and the standard deviation, $\sigma_{\text{\tiny V}}$. In our dry mechanical coupling model, the average speed of agents is conserved under chain formation. Hence, we can estimate $\overline{V}$ directly from the full experimental dataset across all chain lengths, {using the weighted average} $\overline{V} = \sum_{k=1}^N n_k v_k/\sum_{k=1}^N n_k \approx {\SI{9.18}{\micro\meter\per\second}}$, where $n_k$ and $v_k$ {denote the lengths and time-averaged speeds of all active agents detected in experiments, including both chains and solitary bacteria, with a total number of $N=9,525$ data points}. The standard deviation $\sigma_{\text{\tiny V}}$ is the only remaining free parameter, which we determine through a one-parameter fit of the theoretical model evaluated at time $t_*$ to the experimental data in Fig.~\ref{fig:exp}(e). The best fit is obtained for $\sigma_{\text{\tiny V}} \approx {\SI{7.26}{\micro\meter\per\second}}$, and the resulting initial speed distribution is shown as a black solid line in the upper panel of Fig.~\ref{fig:exp}(f). Relative to the mean speed, $\sigma_{\text{\tiny V}}/\overline{V} \approx 0.8$, this fitted value is consistent with previous observations of swimming \textit{E.~coli}.

The excellent agreement between the average chain speed predicted by the numerical simulations and the one observed experimentally, shown in Fig.~\ref{fig:exp}(e), is a direct consequence of our approach to finding $\sigma_{\text{\tiny V}}$. Remarkably, this agreement extends beyond the data used in the fitting procedure. Although it uses only a single fitting parameter to match the average speed of bacterial chains, the theoretical model captures the shape of the full speed distributions remarkably well, as shown in Fig.~\ref{fig:exp}(f). A further indicator of the agreement between theory and experiments is the predicted time scale of aggregation.
The average value of $t_*$ over 50 independent simulations was $\langle t_*\rangle \approx 1.95\,t_0$. Based on the area density of bacteria in the experiments, $c^{2D}_{bact} = 10^3~\SI{ }{\per\milli\meter\squared}$, and the width of the swimming lane given by the bacterial cell body width, $w = \SI{1}{\micro\meter}$, we estimate the linear density as $1/d_0 = wc^{2D}_{bact} = \SI{1}{\per\milli\meter}$ and the characteristic time scale of aggregation as $t_0 = d_0/\overline{V} \approx \SI{1.8}{\minute}$. Thus, the dynamical age of aggregation in our numerical simulations is $t_* \approx \SI{3.5}{\minute}$, {which is of the same order of magnitude as, but not exactly equivalent to, the duration of} the experimental observations, which were collected {between \SI{2}{\minute} and \SI{30}{\minute}} after the onset of aggregation.

The agreement between theory and experiment should be interpreted in light of the simplified nature of the model, which neglects fragmentation, temporal fluctuations in speed, {nematic elasticity}, and detailed hydrodynamic interactions. In particular, several factors may contribute to the fact that our theoretical model underestimates the age of the aggregation process. Incorporating {stochastic reversals of swimming direction into the simulations \cite{Goral2022, Prabhune2024_tugofoars}} would reduce the rate of encounter between active agents, thereby delaying the aggregation process. Including hydrodynamic interactions between bacteria within a chain would likely increase the slope of the speed--length curve at any given time in the simulations, since such interactions would reduce the net viscous drag on the closely packed cell bodies (see Fig.~\ref{fig:exp}(g)) and increase the effectiveness of flagellar thrust in a chain configuration, leading to larger swimming speeds.
As a result, we would expect the simulations to match the experimental measurements at a later time point than is currently estimated. Additional factors, such as the alignment of cell bodies and flagella within a swimming chain, the variability in cell body size and flagellar number, and the {elasticity and} anisotropic viscosity of the LC are also expected to influence the quantitative features of aggregation in ways that are difficult to predict without further modelling.

The theoretical framework developed in this study is intentionally minimal and neglects several physical ingredients that may influence aggregation in real bacterial suspensions, leaving room for several extensions in future theoretical and experimental work. Under the dry mechanical coupling model, our simulations of 1D active aggregation share similarities with a passive 1D ballistic aggregation model \cite{Carnevale1990_ballistic,Piasecki1992_ballistic,Frachebourg1999_1D_ballistic}, which has recently been applied to the coalescence of droplet trains in the limit of unidirectional motion, $P(\text{right})=1$ \cite{Parrenin2024,VaniBonn}. 
In our active aggregation framework, the conservation of momentum, $mv$, is replaced instead by the conservation of viscous drag, $\zeta v$. However, the two systems are not analogous since our model includes active reorientation during head-on collisions, {a distinctive feature of motile microorganisms capable of reorienting their propulsive flagellar structures}. Apart from the initial randomness in bacterial swimming speeds, the simulations are entirely deterministic. In particular, the model does not account for the reversal dynamics of bacteria in nematic LCs \cite{Goral2022,Prabhune2024_tugofoars}, which may delay aggregation by shifting the motion of bacteria {(individuals or chains)} from the ballistic to the diffusive regime on intermediate time scales.

Several features of the experimental protocol also limit the comparison between experiment and theory. First, we do not have direct access to the speed distribution of bacteria before aggregation begins. The speed distribution shown in the upper panel of Fig.~\ref{fig:exp}(f) reflects the behaviour of solitary bacteria left behind after faster swimmers have already formed chains. In addition, the swimming statistics reported in Fig.~\ref{fig:exp}(e--f) combine measurements taken at different times relative to the onset of aggregation, whereas the simulations provide time-resolved predictions from a well-defined initial state. {Due to the preparation protocol for the aligning chamber, a well-defined $t=0$ could not be established in the experiments. }

\section{Conclusion}

In this study, we investigated the aggregation of \textit{E.~coli} bacteria swimming in nematic LCs and showed that their active aggregation dynamics are likely driven by \blue{cell-to-cell variations} in swimming speed, at a rate substantially faster than the elasticity-driven aggregation of passive colloids in LCs. By combining large-scale detection and tracking of bacterial chains with minimal theoretical models built on a parametrized lognormal representation of the speed distribution of individual bacteria, we demonstrated that the observed increase in average chain speed with chain length is likely due to an active self-sorting process in which faster bacteria are preferentially incorporated into longer chains. In particular, our 1D agent-based simulations provide a dynamic picture of a self-organized process in which active agents encounter each other at a rate limited by their relative velocities, hence the term ``swimming-limited aggregation''. 

Altogether, these results establish a quantitative framework for further exploration of the dynamics of swimming-limited aggregation and highlight the crucial role of activity and initial conditions in shaping emergent collective behaviour. {The self-sorting of swimmers driven by \blue{cell-to-cell variations} in speed may confer a biological advantage to bacteria in anisotropic media, such as more efficient collective transport or an earlier colonization strategy.} {Future work could explore the role of nematic elasticity in enhancing and stabilizing motility-induced phase separation in 1D \cite{TailleurCates2008,TailleurCates2015}. Beyond the specific case of bacteria in liquid crystals, the dynamic speed-sorting mechanism identified here may provide a generic route for motility-based self-organization in confined or anisotropic environments, where \blue{cell-to-cell variations} can be converted into spatial and collective heterogeneity through purely physical interactions. }

\section*{Acknowledgements}

{We thank Irmgard Bischofberger and Dong Ki Yoon for assistance in refining the chamber preparation process, and Angela Dawson for developing the dual-colour bacterial strain. We thank Eric Clément, Benjamin Péréz-Estay, Shuang Zhou, Sujit Datta, Daniel Bonn, and Nathan Vani for useful discussions.}

\textbf{Funding:}  A.L., G.S. and M.G. acknowledge funding from the European Research
Council (ERC) Consolidator Grant PaDyFlow (Agreement 682367). {This work has received support from the Institut Pierre-Gilles de Gennes under the Major Research Program of PSL Research University ``IPGG'' launched by PSL Research University and implemented by ANR with the reference ANR-10-IDEX-0001.} {T.L.L., G.S. and M.G. acknowledge funding from the French National Research  Agency,  grant ANR-18-CE09-0028-02. M.T.C. acknowledges support from Clare College, University of Cambridge through a Lynden-Bell Research Fellowship, and from the Institute of Physics, University of Amsterdam through an Els Koffeman Postdoctoral Fellowship.}


\textbf{Codes and Data Availability:} The Python scripts used for image analysis are available at \url{https://github.com/gsintes/Chains}. The MATLAB scripts used for the agent-based simulations and for figure generation, along with the processed data supporting this study, are available at \url{https://github.com/MariaTatuleaCodrean/SLA}. The raw data and experimental videos are available from the authors upon request.

\textbf{Competing Interests:} The authors declare that they have no competing interests.

\bibliography{references}

\newpage

\appendix

\section{Materials and Methods}

\label{app:methods}

\subsection{Bacterial Strains and Culture}

\subsubsection{\textit{E. coli} RP437}

For the majority of tracking experiments, we used the \textit{Escherichia coli} (\textit{E. coli}) strain RP437, a wild-type strain with run-and-tumble motility derived from the K12 strain \cite{Parkinson1978_Ecolimutants}. 
The RP437 strain carries a plasmid expressing yellow fluorescent protein (YFP) that makes the bacterial cell body fluorescent with an excitation wavelength of \SI{513}{\nano\meter} and an emission wavelength of \SI{527}{\nano\meter}, allowing cell body tracking under the microscope.
To prepare the culture solution, we placed \SI{100}{\micro\litre} of frozen stock in \SI{10}{\milli\litre} of M9G culture medium, with \SI{10}{\micro\litre} of chloramphenicol antibiotic to avoid contamination. The culture was incubated overnight at \SI{30}{\degreeCelsius} with shaking at 250 RPM. 
The following morning, the optical density at \SI{600}{\nano\meter} was measured, and cultures with a target OD600 of approximately 0.5 were used for subsequent experiments, as this corresponds to the end of the exponential growth phase.
After OD measurements, the solution was centrifuged at 5000 RPM for \SI{5}{\minute}, rinsed with motility buffer (MB), centrifuged again, and resuspended in MB. The final resuspension in MB determines the final OD of the bacterial solution, which correlates to its bacterial concentration. For an OD600 of 1.0, the concentration is approximately $2.66 \times 10^9$ bact/mL \cite{YapTrau2019_OD600}.

\vspace{-1\baselineskip}

\subsubsection{\textit{E. coli} {AD62}}

\blue{To visualize the ``super-bundle'' configuration of flagella within bacterial chains (Fig.~\ref{fig:exp}(g) and Fig.~\ref{figS:super-bundle})}, we used the \textit{E. coli} strain AD62, which was engineered to express green fluorescent protein (GFP) in the cell body and to display specialized dye-binding receptors on the flagella, allowing for dual-colour staining \cite{Junot2022_AD62}. GFP has an excitation peak at \SI{395}{\nano\meter} and an emission wavelength of \SI{509}{\nano\meter}, while the Alexa fluorescent dye used for flagellar staining is excited at \SI{649}{\nano\meter} and emits at \SI{671}{\nano\meter}.
To prepare the culture solution, the bacteria were inoculated in ampicillin-supplemented LB growth medium, then incubated overnight at \SI{30}{\degreeCelsius} with shaking at 250 RPM. The next morning, 1\% of the bacterial culture was transferred to TB growth medium and incubated for 4--5 hours until it reached a target OD600 of 0.5. 
The bacteria were then centrifuged twice in Berg's motility buffer (BMB) at 5000 RPM for \SI{6}{\minute}, with an intermediate rinse in BMB. Alexa dye (Maleimide Alexa Fluor\textsuperscript{TM} 647 C2 from Thermo Fisher Scientific) was subsequently added to the bacterial solution, and the sample was placed on a rotation plate for 2--3 hours. The bacteria were then washed (centrifuged and resuspended in motility buffer) three times, to remove excess dye, and were finally resuspended for subsequent use. 

\subsection{Liquid Crystal Sample Preparation}

We used the biocompatible liquid crystal Cromolyn sodium salt (DSCG), purchased from Sigma-Aldrich. To preserve bacterial motility, the DSCG molecules were dispersed in MB instead of water. The solution was placed on the vortex
shaker for around \SI{15}{\minute} until fully dissolved, then transferred to an ultrasonic bath for \SI{30}{\minute} to disperse the remaining aggregates.  To prepare a mixture of liquid crystal (LC) and bacteria, we added 20\% or 30\% bacterial solution to a DSCG solution with an initial concentration of 17.1 wt\%, yielding final DSCG concentrations of 13.7 wt\% and 12 wt\%, respectively. These concentrations remained safely above the isotropic--nematic phase transition at a DSCG concentration of 10.5 wt\%, allowing us to retain bacterial motility at room temperature in an anisotropic environment that is not prohibitively viscous \cite{Goral2022}. 
The bacteria were observed in a phosphate MB enhanced with L-serine to maintain motility in oxygen-deprived samples for up to 4 hours. The solution was inserted by capillarity into an aligning chamber for observation under the microscope.

\subsection{Aligning Chamber Preparation}
\vspace{-.35\baselineskip}

The chambers were constructed in a sandwich configuration using either two glass slides or a glass slide and a cover slip, separated by a Parafilm spacer. For experiments using the long working distance 40× objective, the chamber was created by sandwiching two scratched glass slides. When the 63× objective was required for flagellar tracking experiments, we used a mixed configuration consisting of a scratched glass slide and a standard cover slip, which offered enhanced visualization but slightly reduced alignment quality. The assembly was gently heated and pressed together, resulting in a chamber with a controlled thickness of $h = 40 \pm \SI{10}{\micro\meter}$. The chamber was attached to a large cover slip and sealed with wax to prevent evaporation.

To prepare the scratched glass slides, we used a technique based on the method developed by \citet{Suh2018Nanoscratching}. Diamond microparticles were rubbed in a controlled direction to create nano-scratches on the glass surface, providing sites for the LC mesogens to align and establish a preferred orientation in the chamber. This procedure can only be performed on glass slides, since cover slips are too fragile. The glass slides were manually rubbed using a diamond paste from TechDiamondTools and a velvet cloth. The rubbing was performed three times with high pressure along a straight path. The glass slides were then sectioned into small rectangles \SI{2}{\centi\meter} long and \SI{1}{\centi\meter} wide, with the long axis parallel to the rubbing
direction.

\vspace{-.35\baselineskip}
\subsection{Fluorescence Microscopy}
\vspace{-.35\baselineskip}

The experiments were conducted using a Zeiss inverted microscope equipped with a controllable motorized stage and a fluorescent lamp fitted with a cyan colour filter at a wavelength of \SI{511}{\nano\meter}. Images were captured with a Hamamatsu Orca Flash 4.0 camera, using a 40× long working distance objective.
For fixed swimming experiments, images were recorded at 30 frames per second at a fixed position. To visualize the body and flagella simultaneously, {the camera was fitted with a 63× objective}, and a beam splitter was added to the observation microscope. This setup enables blue illumination for the body and red illumination for the flagella.

\vspace{-.35\baselineskip}
\subsection{3D Tracking}
\vspace{-.35\baselineskip}

We used a 3D Lagrangian tracking device developed in the PMMH laboratory by \citet{Darnige2017} \blue{to visualize the ``super-bundle'' configuration of flagella over long periods (Fig.~\ref{fig:exp}(g) and Fig.~\ref{figS:super-bundle}) and to capture rare merging events such as the one depicted in Fig.~\ref{fig:exp}(d)}. The device integrates a 3-axis motorized microscope stage, a Hamamatsu Orca Flash 4.0 camera, and LabVIEW software. The stage is controlled by velocity commands, while real-time image analysis adjusts the stage movement to track the bacteria.
The $(x,y)$-position of the bacterial body is determined directly from the image, while the $z$-position is calculated using a standard technique combining an out-of-focus pattern analysis with a dichotomic search to minimize the size of the detected object \cite{Darnige2017}. Tracking performance was significantly improved in a subsequent version of the software using deep learning methods \cite{Darnige2026}.
For the dual-colour fluorescently stained bacteria, the tracking is performed only on the cell body \cite{Junot2022_AD62}.

\vspace{-.35\baselineskip}
\subsection{Image Analysis}
\vspace{-.35\baselineskip}

We provide a brief description of the image analysis pipeline for detecting and tracking bacterial chains. Further details can be found in the doctoral dissertation of \citet{GuillaumeThesis}.

\vspace{-.35\baselineskip}
\subsubsection{Chain Detection}
\vspace{-.35\baselineskip}

The background signal was calculated as the minimum intensity value of each pixel over the full duration of the experimental video and was subtracted from the original image. This procedure removed bacteria adhered to the glass slide, which appeared as persistent bright marks (high pixel intensity), in contrast to motile bacteria that produced only a temporary increase in pixel intensity and thus were not captured in the background signal. 
To remove out-of-focus bacteria, we applied a Gaussian filter to the resulting image and subtracted the filtered image from the original. This operation preserved only high-contrast objects with sharp boundaries, since out-of-focus features were largely unaffected by the Gaussian filtering and were eliminated during subtraction. The resulting image was binarized, and objects smaller than the typical size of a single bacterium were removed.

{Due to noise, the gap between adjacent bacteria may be detected as either black or white, causing a chain to be segmented into two objects or merged into one, respectively. We empirically observed that this gap was typically 8 pixels. To ensure consistent detection of bacterial chains as single objects, we filled the space between detected objects as follows. For each detected object, a white rectangle was centred on the object centroid, with its width equal to the minor axis and its length equal to the major axis plus 8 pixels.} In this way, we connected neighbouring bacteria within the same chain without increasing its width and inadvertently connecting chains swimming parallel to each other. The objects were redefined based on the newly connected white regions, and ellipses were fitted to each detected object to extract its length, width, centroid, and orientation. The chain length, expressed as a number of bacteria, was determined from the length of the detected object, by subtracting the 8 pixels added during the connection operation and dividing by the typical length of a bacterium.

The final verification step consisted of overlaying the ellipses drawn around the detected objects onto the original image. Most chains were detected accurately, while some marginally out-of-focus bacteria were not detected. Missed detections reduced the number of data points available for statistical analysis, without impacting the qualitative nature of the results. The detection performance was consistent, and random validation checks indicated that the estimated number of bacteria was accurate, with rare errors limited to a deviation of one bacterium.

\subsubsection{Chain Tracking}

Once the bacterial chains were detected as objects in each frame, we used a tracking algorithm to assign consistent ID numbers to the objects as their positions changes over time. The tracking problem can be formulated as an optimization problem, in which we seek to minimize the cost of pairing a tracking ID from frame $t$ with a detected but unassigned object in frame $t+1$. We used the cost function $C_{i,j} = \left( \frac{d_{i,j}}{D_0} \right)^2 + 
          \left( \frac{\delta \theta_{i,j}}{\Theta_0} \right)^2 + 
          \left( \frac{\delta A_{i,j}}{A_0} \right)^2,$
where $d_{i,j}$ is the distance between the new and previously assigned objects, $\delta\theta_{i,j}$ is the difference in orientation, and $\delta A_{i,j}$ is the difference in area. All quantities were rescaled by a factor that sets the accepted variation and relative importance of each tracked feature. Since the expected variation in area was minimal, we chose a small value for $A_0$. Likewise, we expected angular variations to be negligible and adopted a small value for $\Theta_0$, since bacteria were constrained to swim along the nematic director. 

Including the displacement, $d_{i,j}$, in the cost function was effective in low-density systems, where the expected distance travelled by an object between two frames was small relative to the typical separation between objects. 
To avoid misidentification in crowded environments, we implemented a Kalman filter that takes into account the particle's history. The Kalman tracking technique generates ID assignments by comparing the position of detected objects in frame $t+1$ with the predicted future positions of previously identified objects, based on their velocities and positions in frame $t$. {The displacement, $d_{i,j}$, in the cost function was thus redefined as the difference between the predicted and measured positions of objects in frame $t+1$.} After assignment, each particle's position and velocity were updated using a weighted average of the measured and predicted values. 

Examination of the tracking results in the videos suggests that the Kalman tracking algorithm performed well, with accurate assignments even in crowded regions. When the algorithm performed poorly, it was primarily due to the quality of detection before tracking.

\newpage
\section{Estimation of the Passive Aggregation Time Scale}
\label{app:estimation_passive}

To estimate the strength of attractive interactions {induced by the liquid crystal (LC)} and the expected time scale of aggregation for passive, de-flagellated bacteria, we consider the aggregation of two spherical particles of radius $R$ placed in a LC whose properties are captured by only three parameters: a single elastic constant $K$, an effective viscosity $\eta$ along the nematic director, and an anchoring energy $W$ describing the strength of the planar anchoring conditions on the surface of the particles. The particle separation $d$ decreases as $\dot{d}=-2v$, where the speed $v$ of each passive particle arises from a balance between the elastic force,
\begin{equation}
    F_{elast} = C\frac{W^2R^2}{K}\left(\frac{2R}{d}\right)^6,
\end{equation}
with $C$ being a constant of order one \cite{Musevic2018_chains}, and the viscous drag on the particle,
\begin{equation}
    F_{hydro}=6\pi\eta R v.
\end{equation}

These expressions are valid when the particles are sufficiently far apart, so that we may neglect hydrodynamic interactions between the approaching particles and the complex structure of the nematic director in the intervening gap. In this limit, we can rewrite the force balance as a differential equation for the particle separation, $\dot{d}d^6= -\alpha d_0^7$,
with solution $d(t) = d_0(1-7\alpha t)^{1/7}$,
where $d_0$ is the initial particle separation and
\begin{equation}
    \alpha \sim \frac{W^2}{6\pi\eta K}\left(\frac{2R}{d_0}\right)^7
\end{equation}
is a characteristic rate of aggregation. 
This provides an estimate of the time of encounter between two {passive spherical} particles, $t_e \sim (7\alpha)^{-1}$,
as a function of the particle radius, initial separation, and the properties of the LC. Using typical values $\eta = \SI{18}{\milli\pascal\second}$, $K=10$ pN, and $W=10^{-5} ~\text{J m}^{-2}$ for the viscosity of the LC along the director \cite{Habibi2019}, the elastic constant \cite{Zhou2014_Kvalue}, and the anchoring energy \cite{Mushenheim2014}, we estimate that 
\begin{equation}
    t_e \sim \tau_0\left(\frac{d_0}{R}\right)^7,
    \label{eq:app:te_passive}
\end{equation}
where $\tau_0 \approx 6 \times 10^{-7}\SI{ }{\minute}$. 
Note that the anchoring energy $W$ has never been measured directly for Gram-negative \textit{E.~coli} bacteria and its precise value could depend on the chemistry of the membrane, thereby varying between bacterial species.  Here, we use the order of magnitude estimated by Mushenheim \textit{et al.}~for a similar system consisting of rod-shaped, Gram-positive \textit{B.~subtilis} bacteria in a DSCG liquid crystal \cite{Mushenheim2014}. 

From Eq.~\eqref{eq:app:te_passive}, we predict that the encounter time grows rapidly with the distance between particles, as $t_e\sim d_0^7$. Thus, when the particles start closer than ten radii apart, $d_0<10 R$, the encounter occurs within approximately \SI{6}{\minute}. However, the encounter takes longer than \SI{1}{\hour} when $d_0>14R$. This suggests that only neighbouring bacteria could be passively aggregated by the LC on experimentally accessible time scales of less than an hour. 

To estimate the typical distance between bacteria, we need to consider the linear density of bacteria in our samples. The attraction is one-dimensional, and bacteria aggregate only if they are in the same \SI{1}{\micro\meter} wide ``lane'' in the nematic LC (see Supp.~Fig.~\ref{figS:sliding}). By partitioning a \SI{1}{\milli\meter} square into 1000 ``lanes'' of \SI{1}{\micro\meter} width, and using the area density of bacteria in our samples, $c^{2D}_{bact} = 10^3~\SI{ }{\per\milli\meter\squared}$, we derive an estimate for the linear density of $c^{1D}_{bact}= \SI{1}{\per\milli\meter}$. Thus, the average distance between bacteria is on the order of \SI{1}{\milli\meter} or around $300R$. {At such large distances, the passive aggregation time scale for non-motile bacteria vastly exceeds the duration of our experiments. This suggests that the observed aggregation between motile bacteria cannot be explained solely by the attractive interactions induced by the LC, but it is based on an important contribution from active swimming.}

\newpage
\section{Supplementary Figures}

\setcounter{figure}{0}
\renewcommand{\thefigure}{S\arabic{figure}}
\renewcommand{\theHfigure}{S\arabic{figure}}

\begin{figure}[h!]
\centering
\includegraphics[width=\textwidth]{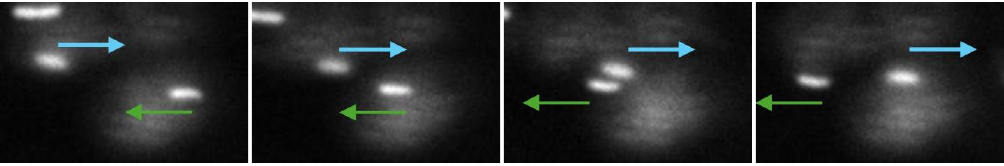}
\caption{Two slightly misaligned bacteria approach head-to-head \blue{(strain:~\textit{E. coli} {RP437})}. They slide past each other without forming a chain, maintaining a gap of approximately \SI{1}{\micro\meter} between them. The images are displayed in temporal sequence from left to right.}
\label{figS:sliding}    
\end{figure} 

\begin{figure}[h!]
\centering
\includegraphics[width=\textwidth]{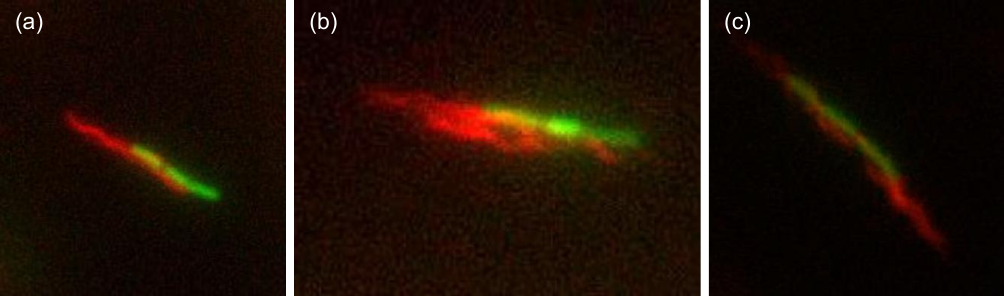}
\caption{Flagella organized in ``super-bundle'' configurations around swimming bacterial chains \blue{(strain: \textit{E. coli} {AD62})}. (a) A chain of three bacteria swimming to the right, with their flagella tightly arranged in a coherent super-bundle. (b) A chain of three bacteria swimming to the right with their flagella arranged in two parallel bundles or a loose super-bundle. (c) A chain of three bacteria swimming to the left with their flagella in a bipolar configuration. A well-organized super-bundle pushes the chain towards the left, while a secondary smaller bundle at the front of the chain, containing at least one flagellum from the leading bacterium, pushes in the opposite direction.}
\label{figS:super-bundle}    
\end{figure} 

\newpage
\section{Nearest-Neighbour Encounter Probabilities}
\label{app:probab}

In this appendix, we use uppercase letters to denote random variables, while lowercase letters are reserved for control variables (e.g., $v_c$) and for specific realizations of random sampling processes. 

\subsection{Relative Velocity Distribution}

The setup of the minimal model of aggregation between three isolated bacteria is depicted in Fig.~\ref{fig:prob}(a). The central bacterium swims forward at a given speed, $v_c$, while the signs $S_{r,f}$ indicate the swimming directions (+1 forward, -1 backward) of its rear and front neighbours, and $V_{r,f}$ indicate their absolute speeds. The apparent velocities of the central bacterium relative to its rear and front neighbours are thus given by $\hat{V}_r = v_c-S_r V_r$ and $\hat{V}_f = v_c - S_f V_f $, respectively. Using the notation of Fig.~\ref{fig:prob}(a), we identify the signed velocities of the peripheral bacteria as $V_{\mathrm{rear}}=S_r V_r$ and $V_{\mathrm{front}}=S_f V_f$.

The absolute speeds of the two neighbouring bacteria are independent and identically distributed random variables with cumulative distribution function $F_V$, such that $p(V_r\leq v) = p(V_f\leq v) = F_V(v)$. Therefore, the total probability that the relative velocity lies below a certain threshold, $\hat{v}$, is given by
\begin{equation}
    p(\hat{V}_k\leq\hat{v})=\sum_{s\in\{-1,+1\}}p(S_k=s)p\left(\hat{V}_k\leq\hat{v}|S_k=s\right),
\end{equation}
where the subscript $k \in \{r,f\}$ may refer to the rear or front bacteria. Furthermore, since bacteria are equally likely to swim left or right (no macroscopic symmetry breaking), we have
\begin{equation}
    p(\hat{V}_k\leq\hat{v})=\frac{1}{2} p\left({V}_k\leq \hat{v} - v_c\right) + \frac{1}{2}p\left({V}_k\geq v_c - \hat{v}\right),
\end{equation}
or, using cumulative distribution function (CDF) notation, simply
\begin{equation}
    p(\hat{V}_k\leq\hat{v}) = \frac{1}{2} + \frac{1}{2}F_V(\hat{v} - v_c) - \frac{1}{2}F_V(v_c - \hat{v}).\label{eq:app:pVhat1}
\end{equation}
Since only one of the latter two terms can be non-zero at the same time, we simplify the result as follows,
\begin{equation}
    p(\hat{V}_k\leq\hat{v}) = \frac{1}{2} + \frac{1}{2}\text{sgn}\left(\hat{v} - v_c\right)F_V(|\hat{v} - v_c|).\label{eq:app:pVhat}
\end{equation}

\subsection{First Encounter Time Distribution}

Having described the probability of observing a certain range of relative velocities, we now consider the expected encounter time between the central bacterium and its nearest neighbours. The central bacterium will encounter its rear or front neighbours only if $\hat{V}_r<0$ or $\hat{V}_f>0$, respectively. The apparent time of these encounter events is $d_0/|\hat{V}_{r,f}|$, with the time of first encounter given by the minimum of the rear and front encounter times, $T_e=\min(T_r, T_f)$. We observe the first encounter within a finite time, $t$, with probability
\begin{equation}
    F_{T_e}(t)=p\left(\min(T_r, T_f)\leq t\right)=1 - p(T_r\geq t)p(T_f\geq t),
\end{equation}
or, using CDF notation, simply
\begin{equation}
    F_{T_e}(t)=F_{T_r}(t)+F_{T_f}(t)-F_{T_r}(t)F_{T_f}(t). \label{eq:app:Te}
\end{equation}
The time of encounter with the rear bacterium is described by
\begin{equation}
    F_{T_r}(t) = p(T_r\leq t)= p\left(\hat{V}_r\leq -\frac{d_0}{t}\right). 
\end{equation}
Using Eq.~\eqref{eq:app:pVhat} and the fact that $v_c>0$ by construction, we obtain
\begin{equation}
    F_{T_r}(t) = \frac{1}{2}-\frac{1}{2}F_V\left(v_c+\frac{d_0}{t}\right).\label{eq:app:Tr}
\end{equation}
Meanwhile, the time of encounter with the front bacterium is described by
\begin{equation}
    F_{T_f}(t) = p(T_f\leq t) = p\left(\hat{V}_f\geq \frac{d_0}{t}\right),
\end{equation}
where we again use Eq.~\eqref{eq:app:pVhat} to obtain
\begin{equation}
    F_{T_f}(t) = \frac{1}{2}+\frac{1}{2}\text{sgn}\left(v_c-\frac{d_0}{t}\right)F_V\left(\left|v_c-\frac{d_0}{t}\right|\right).\label{eq:app:Tf}
\end{equation}
Hence, combining Eqs.~\eqref{eq:app:Te}, \eqref{eq:app:Tr} and \eqref{eq:app:Tf}, we obtain
\begin{equation}
    F_{T_e}(t) = 1 - \frac{1}{4}\left[1+F_V\left(v_c + \frac{d_0}{t}\right)\right]\left[1-\text{sgn}\left(v_c-\frac{d_0}{t}\right)F_V\left(\left|v_c-\frac{d_0}{t}\right|\right)\right].
    \label{eq:app:F_Te}
\end{equation}
In summary, it is possible to reconstruct the probability distribution of $T_e$, the first encounter time in a minimal model of three isolated swimming bacteria, from the cumulative distribution function $F_V$ of the speed distribution of individual bacteria.

\subsection{Infinite Encounter Time Probability}

Naturally, it is also possible that the central bacterium never meets either its front or rear neighbours if $\hat{V}_r>0$ and $\hat{V}_f<0$. The probability of infinite encounter time is given by
\begin{flalign}
    p(T_e=+\infty) &=p\left(\hat{V}_r\geq0\right)p\left(\hat{V}_f\leq0\right),\\
    &=\frac{1}{2}\left[1 - \text{sgn}(-v_c)F_V (v_c)\right] \times \frac{1}{2}\left[ 1 + \text{sgn}(-v_c) F_V (v_c)\right],\\
    &=\frac{1}{4}\left[1-F_V (v_c)^2\right].
\end{flalign}
In particular, the probability of infinite encounter time reaches a maximum value of $1/4$ for stationary bacteria, since $F_V(0)=0$ when $v_c = 0$, and it approaches zero as the speed of the central bacterium increases, since $F_V(v_c)\to 1$ as $v_c \to \infty$. Note also that $p(T_e=+\infty) = \lim_{t\to\infty}F_{T_e}(t)$ as per Eq.~\eqref{eq:app:F_Te}.

\newpage
\section{Dry Mechanical Coupling Model of Bacterial Chains}
\label{app:drymech}

In this appendix, we use bold uppercase letters to denote velocity vectors, $\boldsymbol{V}$, lowercase letters to denote speeds, $v=|\boldsymbol{V}|$, and regular uppercase letters to denote signed velocities along the nematic director, $V = 
\boldsymbol{V}\cdot \boldsymbol{e}_x$. Tildes will generally be used to distinguish between thrust forces, speeds, and velocities of bacterial chains (with tilde) or individual bacteria (without tilde). 

Since experimental observations indicate that both bacterial cell bodies and flagella remain aligned with the director field (see Fig.~\ref{fig:exp}(g) and Supp.~Fig.~\ref{figS:super-bundle}), we only consider the balance of forces acting along the nematic director.

\subsection{Speed Selection for Individual Bacteria}

The velocity of a freely swimming bacterium emerges from the balance of thrust and drag on its body and flagella \cite{Purcell1977}. 
Let us assume that an isolated bacterium generates a total thrust $\boldsymbol{F}$ by rotating its flagella and experiences a net viscous drag $-\zeta\boldsymbol{V}$ from both its body and flagella as it moves forward through a viscous fluid. In our case, the drag coefficient $\zeta$ is proportional to the effective viscosity of the liquid crystal along the nematic director, $\boldsymbol{n} = \boldsymbol{e}_x$. To achieve overall force balance, $\boldsymbol{F} - \zeta\boldsymbol{V} = \boldsymbol{0}$, the velocity must be equal to $\boldsymbol{V} = \frac{1}{\zeta}\boldsymbol{F}$. 

\subsection{Speed Selection for Bacterial Chains}

A swimming bacterial chain is itself a force-free swimmer since no external force is applied to the chain. Therefore, the total thrust generated by all flagella must balance the total viscous drag on the bodies and flagella of all bacteria in the chain, 
\begin{equation}
    \sum_{k=1}^n \boldsymbol{F}_k - \sum_{k=1}^n \zeta_k \boldsymbol{V}_k = \boldsymbol{0}.
\end{equation}
The dry mechanical coupling model acknowledges that bacteria must swim at the same velocity $\boldsymbol{V}_k = V_{\text{chain}} \boldsymbol{e}_x$ to preserve the integrity of the chain, but neglects all hydrodynamic interactions between bacteria. Therefore,
\begin{equation}
    V_{\text{chain}} = \frac{\sum_{k=1}^n \boldsymbol{F}_k \cdot\boldsymbol{e}_x}{\sum_{k=1}^n \zeta_k}.
\end{equation}
In the absence of hydrodynamic interactions, the thrust and drag exerted on each bacterium remain the same as if the bacteria were swimming independently, with $\boldsymbol{F}_k = \zeta_k\boldsymbol{V}_k$ for every $k$.

Although we have observed large \blue{cell-to-cell variations} in the swimming speeds of individual bacteria, Fig.~\ref{fig:exp}(f), we do not know whether such variations arise from differences in the thrust generation capacity of bacteria ($\boldsymbol{F}$ distribution, which could be related to the number of flagella) or their viscous drag coefficient ($\zeta$ distribution, which could be related to the size of the cell body). For simplicity, we assume that all bacteria have the same drag coefficient, $\zeta_k \equiv \zeta$. This gives
\begin{equation}
    V_{\text{chain}} = \sum_{k=1}^n \frac{\boldsymbol{F}_k \cdot\boldsymbol{e}_x}{n\zeta} = \frac{1}{n}\sum_{k=1}^n V_k.
    \label{eq:app:V_mean}
\end{equation}
In particular, when a chain of $n_1$ bacteria swimming with signed velocity $\widetilde{V}_1$ merges with a chain of $n_2 = n - n_1$ bacteria swimming with signed velocity $\widetilde{V}_2$, we expect that
\begin{equation}
    V_{\text{chain}} = \frac{(\widetilde{\boldsymbol{F}}_1 + \widetilde{\boldsymbol{F}}_2)\cdot\boldsymbol{e}_x}{n\zeta} = \frac{n_1\widetilde{V}_1+n_2\widetilde{V}_2}{n},
    \label{eq:app:V_weighted_mean}
\end{equation}
where $\widetilde{\boldsymbol{F}}_1$ and $\widetilde{\boldsymbol{F}}_2$ are the total thrust forces exerted by flagella on the precursor chains, with $\widetilde{\boldsymbol{F}}_j\cdot\boldsymbol{e}_x = \zeta n_j \widetilde{V}_j$ for $j=1,2$ according to Eq.~\eqref{eq:app:V_mean}.

To connect the equations relating the signed velocities of a bacterial chain and its constituent parts, Eqs.~\eqref{eq:app:V_mean}-\eqref{eq:app:V_weighted_mean}, to the main text equations relating their swimming speeds, Eqs.~\eqref{eq:vchain}-\eqref{eq:vchain_n1n2}, we must consider how the swimming direction changes after an encounter between two swimming agents (either single bacteria or chains).

\subsection{Selection of Swimming Direction}

First, we note the distinction between rear-end and head-on collisions depicted in Fig.~\ref{fig:simu}(a). In rear-end collisions, the  precursor agents swim in the same direction with $\sgn(\widetilde{V}_1) = \sgn(\widetilde{V}_2)$, and Eqs.~\eqref{eq:app:V_mean}-\eqref{eq:app:V_weighted_mean} trivially reduce to Eqs.~\eqref{eq:vchain}-\eqref{eq:vchain_n1n2}. In head-on collisions, we postulate that the new swimming direction is determined through a brief tug-of-war in which the strongest agent prevails.

Our main assumption is that the swimming direction is set by the orientation of the flagellar bundle, and that \textit{E. coli} bacteria can rearrange their flagella after forming a chain due to the flexible flagellar hooks. Immediately after a merging event between a chain of $n_1$ bacteria and signed velocity $\widetilde{V}_1$, and a chain of $n_2$ bacteria and signed velocity $\widetilde{V}_2$, while the flagella remain in their original configuration, the direction of swimming is given by the sign of the net thrust exerted by the flagella on the chain. In other words,
\begin{equation}
    \sgn(V_{\text{chain}}) = \sgn\left[(\widetilde{\boldsymbol{F}}_1 + \widetilde{\boldsymbol{F}}_2)\cdot\boldsymbol{e}_x\right] = \sgn(n_1\widetilde{V}_1+n_2\widetilde{V}_2),
\end{equation}
according to Eq.~\eqref{eq:app:V_weighted_mean}. The stronger agent wins the tug-of-war since the sign of $(\widetilde{\boldsymbol{F}}_1 + \widetilde{\boldsymbol{F}}_2)\cdot\boldsymbol{e}_x$ is equal to the sign of the largest force. 

As the new chain starts to move in this direction, the presence of some flagella at the front of the moving chain represents an unstable configuration. Due to the flexible joints known as flagellar hooks, the flagella can reorient to align with the new direction of swimming \cite{Riley2018Elastohydrodynamic}, so that both sets of flagella contribute a thrust $|\widetilde{\boldsymbol{F}}_1| =\zeta n_1|\widetilde{V}_1|$ and $|\widetilde{\boldsymbol{F}}_2| =\zeta n_2|\widetilde{V}_2|$ in the new direction of swimming. After this flagellar rearrangement, the signed velocity becomes
\begin{equation}
    V_{\text{chain}} = \mathrm{sgn}(n_1\widetilde{V}_1+n_2\widetilde{V}_2)\frac{n_1|\widetilde{V}_1|+n_2|\widetilde{V}_2|}{n},
    \label{eq:app:Vchain}
\end{equation}
and the corresponding swimming speed is
\begin{equation}
    v_{\text{chain}} = \frac{n_1\tilde{v}_1+n_2\tilde{v}_2}{n}.
    \label{eq:app:vchain_n1n2}
\end{equation}
In our agent-based simulations, we assume that the flagellar rearrangement is instantaneous and that newly formed chains obey Eq.~\eqref{eq:app:Vchain} from the moment of aggregation.

By recursively expressing the speeds of bacterial chains in terms of the speeds of their constituent parts, via Eq.~\eqref{eq:app:vchain_n1n2}, the speed of a given chain can be traced back to the intrinsic speeds of individual bacteria in its composition, so that
\begin{equation}
    v_{\text{chain}} = \frac{1}{n}\sum_{k=1}^n v_k.
    \label{eq:app:vchain}
\end{equation}
for a chain of $n$ bacteria.

\end{document}